%% file: main.tex
\documentclass{article}

\usepackage{arxiv}

\usepackage[utf8]{inputenc} 
\usepackage[T1]{fontenc}    
\usepackage{hyperref}       
\usepackage{url}            
\usepackage{booktabs}       
\usepackage{amsfonts}       
\usepackage{nicefrac}       
\usepackage{microtype}      
\usepackage{lipsum}         
\usepackage{graphicx}
\usepackage{soul, color}
\usepackage[citestyle=numeric,backend=biber]{biblatex}
\usepackage{doi}

\usepackage{xcolor,colortbl}
\usepackage{amsmath}
\usepackage{cleveref}       

\usepackage[singlelinecheck=false]{caption}
\usepackage{listings}
\usepackage{subfig}
\usepackage{float}

\usepackage{regexpatch}

\definecolor{mGray}{rgb}{0.5,0.5,0.5}

\lstdefinestyle{CStyle}{
    basicstyle=\small\fontsize{7pt}{8pt},
    numberstyle=\tiny\color{mGray},
    numbers=left,
    numbersep=5pt,
    language=C
}

\PassOptionsToPackage{hyphens}{url}\usepackage{hyperref}

\title{Improving AFL++ CmpLog: Tackling the bottlenecks}

\author{
	Sander J. Wiebing \\
	Student \\
	Vrije Universiteit Amsterdam\\
	Amsterdam, Netherlands, \\
	\texttt{sander@swiebing.com} \\
	\And
  Thomas Rooijakkers \\
	Cyber Security Technologies\\
	TNO\\
	The Hague, Netherlands, \\
	\texttt{thomas.rooijakkers@tno.nl} \\
	\And
	Sebastiaan Tesink \\
	Cyber Security Technologies\ \\
	TNO\\
	The Hague, Netherlands, \\
	\texttt{sebastiaan.tesink@tno.nl} \\
}

\renewcommand{\shorttitle}{Improving CmpLog AFL++}

\hypersetup{
pdftitle={Improving AFL++ CmpLog: Tackling the bottlenecks},
pdfauthor={Sander Wiebing, Thomas Rooijakkers, Sebastiaan Tesink},
pdfkeywords={AFL++, fuzzing, CmpLog, RedQueen},
}

\begin{document}
\maketitle

\begin{abstract}
The performance of the AFL++ CmpLog feature varies considerably for specific programs under test (PUTs). In this paper it is demonstrated that the main cause of the poor performance is low seed entropy, and a lack of deduplication of magic bytes candidates. An improvement is proposed by mapping comparisons to input bytes, in order to track which comparisons are controlled by what input bytes. This mapping is then used to fuzz only the comparison values that are magic byte candidates for that input part. Second, a caching mechanism is introduced to reduce the number of redundant executions. The evaluation of the improved versions shows a significant coverage gain compared to the original AFL++ implementation of CmpLog for all PUTs, without breaking functionality. The proposed solution in this paper provides a solid basis for a redesign of CmpLog.
\end{abstract}

\keywords{AFL++, fuzzing, CmpLog, RedQueen}

\section{Introduction}

Fuzzing, or fuzz-testing, is a software testing technique to find bugs and vulnerabilities in programs by feeding a program under test (PUT) with (randomly) mutated input. Over the past decades, many different fuzzers \cite{manes2019art} and fuzzing techniques have been developed to fuzz faster, increase code coverage, or bypass specific challenges in a PUT. Fioraldi et al. took several fuzzing features from previous research and combined them into AFL++ \cite{fioraldi2020afl++}, an open source community-driven fuzzer based on AFL\cite{zalewski2017american}. AFL++ enabled researchers to easily integrate their research into a single fuzzer, while making it possible to fuzz binaries with a customized configuration by switching fuzzing features on or off. As a result, however, the introduction of some of these features have lead to  a strongly varying performance. This means that the best performing configuration differs for per PUT \cite{chen2019enfuzz}, since certain characteristics of the PUT may require different features enabled for the optimal fuzzer configuration. In this paper, the focus will be on the CmpLog feature of AFL++, since benchmarks indicate that this feature is promising for most binaries, but we demonstrate that it can result in a dramatic drop in performance for others.  Based on this research, either the implementation of Cmplog could be improved, or Cmplog could be disabled when poor performance is predicted based on characteristics of the PUT.

\section{Background: bypassing magic bytes}

In order for the fuzzer to find and trigger a bug, the vulnerable code needs to be reached. Software binaries often contain checks validating whether parts of the input are equal to a preset byte ranges. Those heavily constrained input bytes are often referred to as \emph{magic bytes}. The probability of discovering these magic bytes via random mutations is very low. Therefore, these magic bytes form a strong roadblock for fuzzers while attempting to gain deeper code coverage, and are one of the primary reasons why fuzzers fail to reach and discover deeply nested bugs and vulnerabilities.

The authors of RedQueen \cite{aschermann2019redqueen} observed that many of these `magic byte' checks, are directly correlated with the input. Based on this observation, RedQueen developed an inexpensive technique to detect magic bytes and move them to the input at the right location, a technique refered to as \textit{Input-To-State \(I2S\)}. Based on the improvements demonstrated by RedQueen, CmpLog was implemented in AFL++. CmpLog further boosts the performance of RedQueen by introducing a shared table containing the operand of the last $256$ executions of every comparison. Subsequently, during the colourization stage, CmpLog colourizes the input; it replaces random bytes with random values. This colourization process enables Cmplog to identify I2S comparisons via the shared table and replaces them in the input. Additionally, CmpLog contains \textit{transformations} and \textit{arithmetic} features. When enabled, there are also different transformations and arithmetic operations performed on the input and replaced. This method tries to bypass I2S comparisons in case the input is slightly modified before the comparison takes place.

A more accurate, but very expensive, method for bypassing magic bytes is by using \emph{taint analysis}. Taint analysis tracks the input through the entire program and therefore knows which operands are performed on the input bytes and whether they are compared to other bytes. VUzzer and Angora use this approach \cite{rawat2017vuzzer, chen2018angora}. Although this technique works well, especially for binaries with a lot of hard comparisons, it also causes a significant drop of executions per second compared to mutation-based fuzzers.

Symbolic execution is a third approach for tackling the magic byte challenge. With symbolic execution, the program is executed abstractly by building constraints on the input to reach a certain part of the binary. Symbolic execution is a great way to bypass the magic bytes when the applications are not that complex. However, state explosion is a big problem in larger binaries, while keeping track of all possible paths. Hence it is not feasible to fuzz purely with symbolic execution. Concolic execution is a twist which is better scalable since it only takes one path and can concretize some variables in the process. Several fuzzing approaches use this method to compute magic bytes \cite{stephens2016driller,yun2018qsym,borzacchiello2021fuzzolic}. While Driller \cite{stephens2016driller} switches to concolic when the mutation-based fuzzer gets `stuck', more recent research proposes to run the concolic execution engine in parallel with the fuzzer \cite{yun2018qsym,borzacchiello2021fuzzolic}.

Since both taint analysis and symbolic execution are expensive fuzzing techniques with regard to the number of executions being performed and memory usage respectively, the focus was drawn to the implementation of CmpLog, which will be explored in more detail.

\section{Exploration}

The authors of \cite{chen2019enfuzz} demonstrated that different fuzzer configurations are optimal for different binaries. For benchmarking, \emph{FuzzBench: an open fuzzer benchmarking platform and service}\cite{fuzzbench} is used througout this paper. The binaries with the most deviating performance\footnote{\href{https://www.fuzzbench.com/reports/2022-04-19/index.html}{https://www.fuzzbench.com/reports/2022-04-19/index.html}\label{benchmarkgoogle}} are included in Table \ref{tab:fuzzbench_all}. While analysing these results, two binaries stood out, namely the \textit{bloaty\_fuzz\_target} and \textit{libpcap\_fuzz\_both} target binaries.

AFL++ performs relatively well on the \textit{libpcap\_fuzz\_both} target, just like libFuzzer \cite{serebryany2015libfuzzer}. \textit{libpcap\_fuzz\_both} probably contains a lot of `magic byte' -checks; as the default AFL++ configuration has CmpLog enabled and libFuzzer contains a similar method to bypass those bytes. The AFL++ paper \cite{chen2019enfuzz} confirms this observation; the AFL++ configuration where CmpLog is enabled outperforms all others. The test with the \textit{bloaty\_fuzz\_target} demonstrates inverted results; AFL++ with CmpLog enabled shows a significant drop in performance.

The FuzzBench repository also contains an AFL++ optimal configuration for the various benchmark targets\footnote{\label{optimalgithub}\href{https://github.com/google/fuzzbench/blob/master/fuzzers/aflplusplus_optimal/fuzzer.py}{https://github.com/google/fuzzbench/blob/ 93d4182f1d5f420cd07224d727288ea4107beeee/fuzzers/aflplusplus\_optimal/ fuzzer.py}}. These hand-crafted configurations are manually determined though on a trial and error basis. During the experiments it was observed that, although the total coverage were comparable at the end of the benchmarks (24h), there was a significant difference in the pace of reaching this coverage, as illustrated by the graphs\textsuperscript{\ref{benchmarkgoogle}}. For the \textit{bloaty\_fuzz\_target}, however, there was a large difference; the optimal configuration, without CmpLog and MOpt enabled, strongly outperformed AFL++ default configuration. These results are in line with the earlier experiment of the original AFL++ paper \cite{fioraldi2020afl++}, where both AFL++ with CmpLog and AFL++ with CmpLog and MOpt performed poorly, but AFL++, both without CmpLog, relatively well. MOpt\cite{lyu2019mopt} is a mutation scheduling technique which gives probabilities to the mutation operators for finding a new path. Therefor, it is likely that the performance hit in \textit{bloaty\_fuzz\_target} is caused by the CmpLog functionality.

\input{tables/fuzzbench_all.tex}

\input{tables/optimal.tex}

\section{Analysis}
Based on the previous section, it was hypothesized that the CmpLog feature of AFL++ was the cause of lower code coverage on the \textit{bloaty\_fuzz\_target}, when compared to other AFL++ configurations. In this section, a root cause analysis of this performance drop is performed through the fuzzing results of `AFLplusplus' (default configuration) and `aflplusplus\_optimal' from the fuzzbench 2022-04-22-aflpp experiment\textsuperscript{\ref{benchmarkgoogle}}.

\textbf{Bloaty} --- The bloaty binary is a size profiler for binaries. It is able to compute the size of different data sources (segments, sections, symbols, compile-units, inlines and ar-members) and supports three binary formats (ELF, Macho-O, and WebAssembly).

\subsection{Code coverage}
As a starting point, the coverage mapping of the two configurations is compared. The branch coverage, excluding shared libraries, is included in Table \ref{tab:coverage}. At first sight it looks like AFL++ performs just as good as the optimal variant for all ELF binaries, the biggest coverage difference is observed for binaries in the Macho-O file format and to a lesser extent the WebAssembly binaries. The demangle code is used by these three binary handlers, so its low coverage could be a result of low coverage on the binary handlers.
Through an investigation of the individual not-covered branches of the Macho-O handler, there appeared to be no comparisons that could be disruptive or impossible to solve for CmpLog.

Furthermore, the coverage report demonstrated that the coverage difference between the ELF handler and Macho-O handler was mainly caused by the set of initial seeds. Out of the 94 initial seeds, only a single seed was a Macho-O formatted file. Thus it cannot be concluded that CmpLog performs properly on ELF, since the initial seed already bypasses the magic bytes.

\input{tables/coverage.tex}

\subsection{Statistics}

Since the coverage mapping did not provide a clear indication of why the CmpLog reaches a relatively low coverage, the poor performance may be caused by `slow' performance of the CmpLog feature. To validate this hypothesis some statistics of the trials were produced. Looking at the average statistics of the trials, Table \ref{tab:stats_default_optimal}, it stands out that AFL++ did not complete any cycle while the optimal version already completed about 31. AFL++ completes a cycle when it reaches the end of the seed queue. Newly found seeds during the fuzzing process are added at the back of the queue, so they can be fuzzed during the same round. Note that not always all queue items are fuzzed, more information about the scheduling can be found in the AFL++ documentation \cite{aflplusplus}. While the execution speed of the optimal configuration is almost double, this cannot be the sole cause; in that scenario one would expect AFL++'s default configuration to have completed at least a few cycles. It seems that the CmpLog feature is generating more executions per seed.

\input{tables/stats_default_optimal.tex}

Table \ref{tab:seed_orig} displays the maximum and average time and number of executions spent per seed, ranked by time spent. This confirms that CmpLog generates a lot more executions per seed; on average the max time spent on a single seed is exceeds 3 hours. The average time for the top 15 seeds with CmpLog enabled is 26 times higher. This does explain why AFL++'s default configuration did not complete any cycle in 24 hours. It is important to note that not all seeds take that long, some seeds are processed in less than a minute. 

\input{tables/seed_orig.tex}

To investigate the cause of this `execution explosion', one of the slowest seeds was taken and analysed manually, in order to determine which part of the bloaty target contributed to this explosion. It turned out that the function `cs\_disasm\_iter()', from the shared library Capstone\cite{capstone} which disassembled the binary, had a large share in the cause of the creation of many of the executions. Especially the print function (printInstruction()) was a large contributor to the execution explosion as it contained a huge switch statement to find the corresponding ASCII character for the current instruction. The seeds that took at least 1h to be processed are all ELF binaries, since Bloaty currently only supports disassembling for ELF, and are at least 10kb in size.

\subsection{Implementation}
To gain a better insight into the execution explosion,  the code of the CmpLog implementation was analysed in more detail. Listing \ref{lst:fuzz-loop} shows a simplification of the CmpLog code with the loops as a basis. The first step of the CmpLog state is the colourization phase. In the colourization, every byte of the input is replaced by a different byte of the same type, a series of bytes which are replaced is called a taint region. Initially, the complete input is taken as one taint region, this colourized input is passed to the target program. When the resulting hash of the execution path differs from the hash of the original execution path, the taint region is cut in half and processed separately, otherwise the region is saved. The result after the colourization is a colourized input where the bytes in every taint region are replaced by a different byte of the same type, and the hash of the execution path is equal to the original input. For the construction of the hash, the `trace\_bits' and `map\_size' are used.

Two types of comparisons exist; the immediate values (INS) and the values referenced by a pointer (RTN). Since it is not possible to know the length of the last type, a length of $31$ is copied to the comparison log by default. For INS the last $32$ hits are logged, for RTN only the last $8$ hits. The handling of the types INS and RTN are similar, for simplification only the RTN type is discussed here. 

CmpLog runs the target with the original input and the colourized input and saves both shared tables (containing the logged comparisons). The next step is to loop over every logged hit in each comparison (see \ref{lst:fuzz-loop}, line 16), and within that loop over every taint byte. In this nested loop, the function rtn\_extend\_enconding() is called twice (line 24, 28). During the first evaluation it is assumed that the right operator is the value controlled by the input (pattern) and the left the magic byte (replication). During the second evaluation, the role of the operators is swiched.

The rtn\_extend\_enconding() function replaces the first byte of the replication value and executes the target with this modified input. This will continue until the last byte of the taint region length is reached and as long as the pattern, the value of the other side of the comparison, is equal to the byte in the input which will be replaced. The idea behind this is that if the operator value has an I2S relation with that input byte, the value should be equal. For the INS comparisons this has to hold both for the original input and the colourized input. For the RTN type this only has to hold for either of them. Note that this I2S check is only performed when transformations are disabled, since this check no longer holds when transformations are assumed between the input and the comparisons. With transformations enabled, the target is always executed for the complete taint\_len.
To summarize, the equation below computes the number of executions executed for one seed,
\begin{equation*}
    \text{Executions per seed } = \text{ total Loggeds } * \text{ total taint\_bytes } * (\text{taint\_len where pattern}[i] == \text{ buf}[i]).
\end{equation*}

\subsubsection{Seed \$27\$}

A more in-depth analysis of seed $27$, is performed, one of the seeds susceptible to the execution explosion problem. By disabling the disassemble function and enabling it again, the number of INS comparisons increases by almost a factor $3$ ($282 \rightarrow 743$), while the number of taint bytes stays roughly equal (around $10.1$k out of the $12.6$k input bytes). These two factors seem to have a huge effect on the number of executions since for every logged comparison hit CmpLog loops through every taint byte. The observation that the number of taint bytes does not decrease can be explained by the characteristics of the disassemble function. It contains a huge switch statement at which every byte is translated to an ASCII character to represent assembly. Changing these bytes will result in different characters but will not change the execution path. However, since for \textit{bloaty\_fuzz\_target} the transformations are not enabled, the condition that the pattern had to be equal to the input before it is executed should reduce the number of executions if the entropy of the pattern is high enough.

\begin{lstlisting} [caption={Simplified CmpLog code},label={lst:fuzz-loop},style=CStyle]

  void rtn_extend_encoding(idx) {
    for (i = 0; i < taint_len; ++i) {
      if ((pattern[i] != buf[idx + i] && 
        o_pattern[i] != orig_buf[idx + i])) {
        break;
      }
  
      buf[idx + i] = repl[i];
      its_fuzz(buf);
    }
  }
  
  void rtn_fuzz(CMP) {
  
    for (loggeds in CMP) {
      for (taint in taints) {
        for (idx = taint.start; idx++; 
          idx <= taint.end){
          taint_len = taint.end - idx;
  
          // Assume right-operatator is 
          // the MAGIC byte
          rtn_extend_encoding(idx, ...);
          
          // Assume left-operatator 
          // is the MAGIC byte
          rtn_extend_encoding(idx, ...);
        }
      }
    }
  }
  
  void input_to_state(afl, orig_input) {
    colorized_input = colorization(orig_input);
  
    common_fuzz(orig_input);
    orig_cmp_map = afl.cmp_map;
  
    common_fuzz(colorized_input);
    color_cmp_map = afl.cmp_map;
  
    for (CMP in  CMP_MAP_W) {
      if (CMP.hits == 0) continue;
      if (CMP.type == INS)
        cmp_fuzz(CMP);
      else if (CMP.type == RTN)
        rtn_fuzz(CMP);
    }
  }
  \end{lstlisting}

\subsection{Patterns entropy}
\label{sec:patterns}

The number of executions is influenced by the number of logged comparisons multiplied by the number of taint bytes. Nevertheless, except for transformations, an input where the pattern is replaced by the replication is only executed when the pattern was equal to the input at that location. Thus, as long as the input or pattern has a high entropy, there are only a few places where the replication can be inserted. To understand the casue for the high number of executions, the number of times the target is executed whenever it got logged was tracked.

Table \ref{tab:rtn-operand} shows three examples of RTN (pointer) operands, only the left or right operand is included since the replicated value has no influence on the execution of the input. The rtn\_extend\_encoding() replaces byte by byte the input with the replication value and executes it subsequently until the condition is no longer met; if the byte in the colourization pattern and original pattern differ from the byte to be replaced in the colourized input and original pattern. 
The two operands of example \#1 produced only $28$ executions, there were only a few places where one or more bytes could be inserted. This is not unexpected since the starting bytes seem to have a lot of entropy. It is different for examples \#2 and \#3. Both operands of \#2 start with $8$ zero bytes, resulting in more than $42$k executions. This is more than the input length ($12.6$k) because for each input location, the replicated value is inserted byte by byte until the replicated value is completely inserted or until the condition no longer holds. Note that since it is an \texttt{OR} statement, either the original input or the colourized input has to contain a $0$. The operands of example \#3 start with $24$ zero bytes and produce over $84$k executions; this is only one-half of one logged hit of a comparison. If there are too many of these cases this leads to an execution explosion.

\input{tables/rtn_operand.tex}
\input{tables/ins_operand.tex}

\begin{figure}
  \includegraphics[width = 1\textwidth]{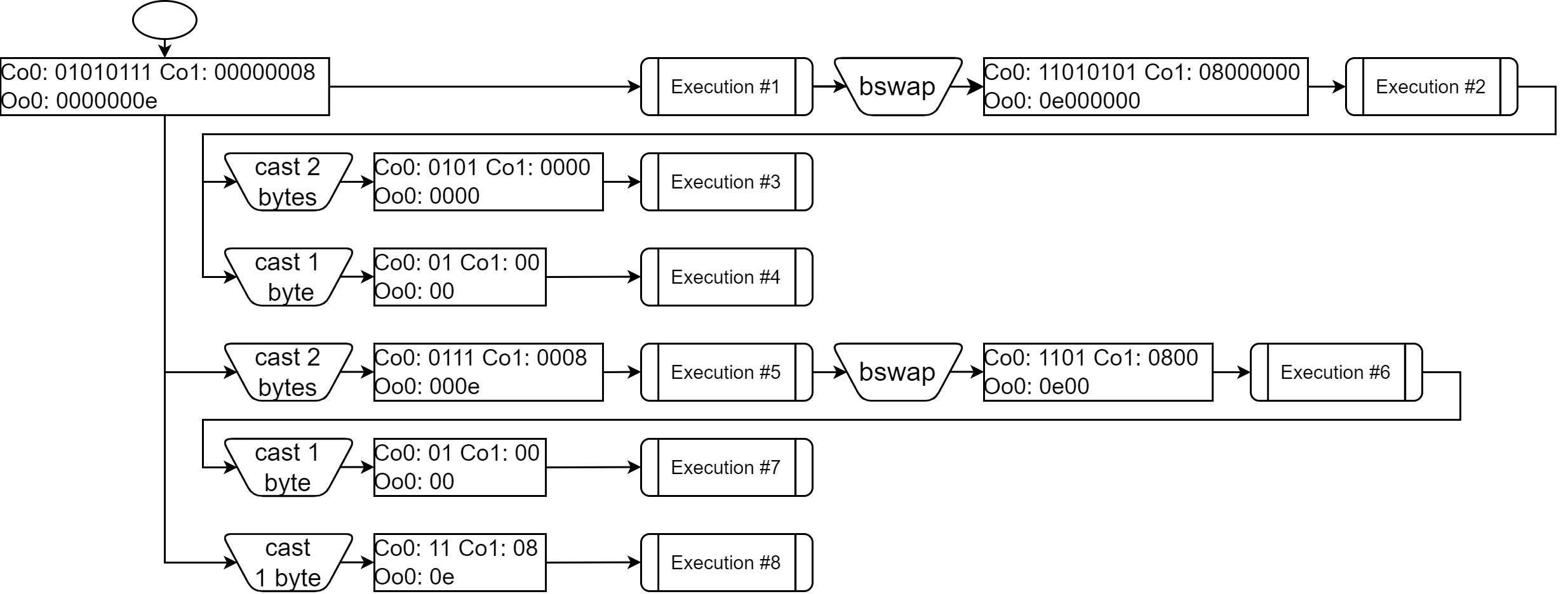}
  \centering
  \caption{The cast and swap flow of Table \ref{tab:ins-operand} example \#3, Co0 = Colourized o0, Co1 = Colourized o1, Oo0 = Original o0.}
  \label{fig:inscastswap}
\end{figure}

The other type of comparison is the immediate values (INS), four examples of operands with the number of produced executions are shown in Table \ref{tab:ins-operand}. The direct values of the immediate operands are not causing that many executions, it is the swapped variants of the operands that do. The variants are generated after an immediate item is fuzzed by swapping the operands and calling cmp\_extend\_enconding() again. 

Figure \ref{fig:inscastswap} demonstrates the cast and swap process and to which executions this operand value may lead. In this setting, Colourized o0 (Co0) and Original o0 (Oo0) are the patterns and Colourized o1 (Co1) is the replication value; the replication value is the value which replaces Co1 in the colourized input. The shown possible executions are only executed if Co0 equals the colourized input and Oo0 equals the original input (the I2S relation). Table \ref{tab:ins-operand} shows in parentheses the result of the "cast 2 bytes" operation followed by "bswap" for each example, execution \#6 in Figure \ref{fig:inscastswap}.

A lot of executions are performed when either the colourized pattern equals `01' (or a series of `01' bytes) and the original pattern equals `00' (or a series of `00' bytes), or vice versa. This may be caused by the behaviour of the colourization phase in combination with the swapping and casting. Namely, the colourization process replaces all `00' bytes in the original input to `01' bytes in the colourized input. So if the colourized pattern equals `01' and the original pattern `00', it will match all locations where the original input is `00'. This becomes a problem when the original input contains a lot of zero bytes which will trigger a lot of executions.

For \#1, Table \ref{tab:ins-operand}, this is not the case, for the 0 side `01' and `06' are observed, for the 1 side `00' and `06'; there are zero locations in the input where these cast patterns match with both the original and colourized input. While examining example \#3 and Figure \ref{fig:inscastswap}, three possible executions were observed matching these `01' and `00' series: \#3, \#4, and \#7. So the I2S relation for these values are met if the original input contains 0, after which the replication value is replaced in the colourized input and executed. For this example more than 20k executions were performed. Note that for all three executions the replication value equals `00', so two executions are duplicates and thus redundant.

For example \#2 the same pattern was observed to a lesser extent, both sides contain the `01' and `00' but the I2S relation will only hold when the result of bswap is cast to 1 byte. This resulted in 7k executions.

Example \#4 results in 13 possible executions after the swap and cast operations, of which 6 of them contain the `01' and `00' series and thereby meet the I2S relation for `0' bytes in the original input. This specific example results in more than $38$k executions.

\subsection{Conclusion}

There are several factors which contribute to the decreased performance of CmpLog on Bloaty fuzz. Firstly, the relatively high number of logged comparisons in combination with a high number of taint bytes. Secondly, the very low entropy of the RTN comparisons and the resulting low entropy due to the swapping of the INS comparisons. In addition, all `0' bytes in the original input are all replaced by `1' bytes in the colourized input, and vice-versa: the colourization of `00' and `01' bytes does not create any entropy. Due to this low entropy in the input and operand, in combination with that for every logged comparison all the taint regions are considered `candidates', there is an execution explosion. 

\section{Improvements}
\label{sec:improvements}

In the previous section, some serious performance issues of CmpLog were revealed for a class of complex PUTs. These performance drawbacks and their impact can be massively reduced as shown in the following sections.

\subsection{Comparison to taint mapping}
The first improvement is to estimate a possible relation between a taint region and a comparison. By utilising knowledge of what comparisons are controlled by which taint region, the only comparisons which are under control have to be replaced by magic bytes and executed subsequently. Both a coarse-grained and a fine-grained approach were considered.

\subsubsection{Coarse-grained}
\label{sec:coarse}
In the coarse-grained approach, the target is initially executed with the original input, followed by an execution with the colourized input. The following step is to compare each logged item of the comparisons. When there is a difference between the two comparisons logs for one or both operands, the operand which is changed in that currently logged item is marked. Next, the original fuzz cycle is entered. Comparisons in which none of the logged compares changed are skipped immediately. In the other comparisons only the changed operands of the logged items are considered magic bytes candidates. This approach enables us to ignore a lot of comparisons in advance and even ignore one of the two operands when they are static.

\subsubsection{Fine-grained}
\label{sec:fine}
The fine-grained approach takes it up a notch. After the first estimation is done, just like in the coarse-grained approach, the taint regions are tested one by one. This is achieved by taking the original input and only replacing the colourized bytes in that taint region. Next, the target is executed again and the comparison logs are compared. For every operand in each logged comparison, it is tracked which taint regions modify the value. So in the fuzzing cycle, only the input bytes in the taint regions related to that operand are checked to be replaced with the replication value and possibly fuzzed.

\subsection{Caching}
\label{sec:caching}

The current design of CmpLog makes the repeated execution of the same input inevitable. The fuzzing loop, as shown in Listing \ref{lst:fuzz-loop}, executes every replication candidate for every comparison regardless of the fact whether the candidate was executed at an earlier stage already. Even worse - as discussed in Section \ref{sec:patterns} - the immediate comparisons produce duplicate inputs for the same operand due to casting and swapping, resulting in the same operand value.

One approach would be to first gather the (to be executed) inputs and perform deduplication before executing them. However, the number of potential inputs for a single seed can grow enourmously, leading to high memory usage. A more robust approach would be to implement a least-recently-used (LRU) caching mechanism to track of the previously replaced values for all input bytes. This mechanism prevents repeated replacement of input bytes and subsequent executions.

The manual inspection of the pattern's entropy - Section \ref{sec:patterns} - also showed that the replication values which are replaced most are the `00' `01' bytes and the bytes of length $2$, being `0000' and `0001'. To demonstrate the potential of implementing caching, an array in CmpLog was imlemented to track for every input byte whether the above-mentioned bytes have been replaced. In the fuzzer loop, it checks whether the ``to be replaced value'' equals one of those four bytes. When this check evaluates to true, it validates whether that byte was replaced before and runs the target or skips the execution.

\subsection{Additional optimisations}
\label{sec:additional}

Originally, the insertion and fuzz loop is only aborted for the current byte $i$, if the following holds

\begin{equation}
  \label{eq:condition}
  \operatorname{color\_pattern}[i]\ != \operatorname{color\_input}[i]\ \&\& \operatorname{orig\_pattern}[i]\ != \operatorname{orig\_input}[i].
\end{equation}

More specifically, it only aborts the loop if both the colourized and original pattern do not match the corresponding input.
However, to meet the I2S relation both the colourized and the original pattern should match with the corresponding input.
The current condition (\ref{eq:condition}) results in a substantial bunch of executions which are not a valid candidate for an I2S relation. For example, consider `01000100' for the colourized operand and `00010001' for the original operand, with input `00000000'. The original statement will replace all four bytes, while it is trivially not an I2S relation. The AND ($\&\&$) with an OR ($||$) were replaced to mitigate this.

\begin{equation*}
  \operatorname{color\_pattern}[i]\ != \operatorname{color\_input}[i]\ || \operatorname{orig\_pattern}[i]\ != \operatorname{orig\_input}[i].
\end{equation*}

Secondly, through manually inspection it was discovered that, especially for the immediate values, quite often a 0-byte was ``replaced'' by a 0-byte and executed subsequently. To mitigate this, a check was implemented that validates whether the byte replacement equals the byte to be replaced for both the pointer comparisons as well as the immediate comparisons, thus skipping these redundant executions. 

\begin{figure*}
  \subfloat[Mean code coverage growth on bloaty\_fuzz\_target]{\includegraphics[width = 0.6\textwidth]{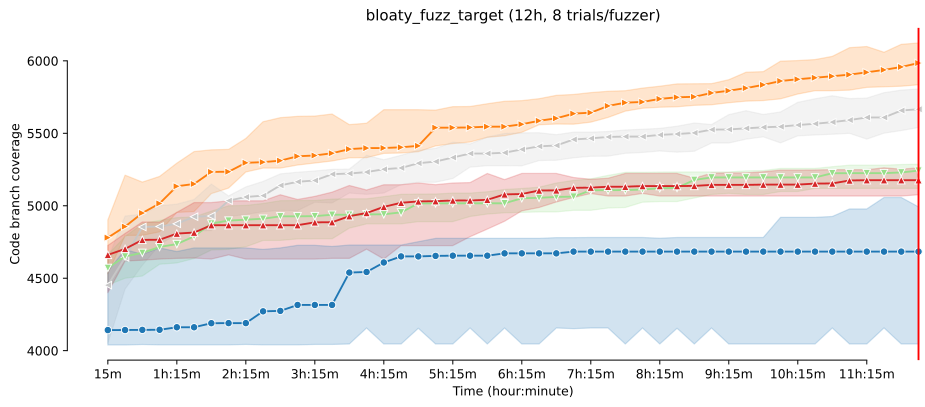}}
  \subfloat[Reached code coverage distribution
  on bloaty\_fuzz\_target]{\includegraphics[width = 0.3\textwidth]{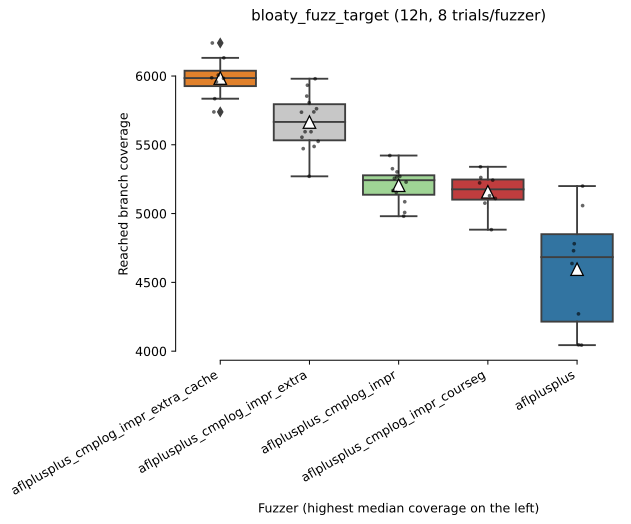}} \\

  \subfloat[Mean code coverage growth on libpcap\_fuzz\_both]{\includegraphics[width = 0.6\textwidth]{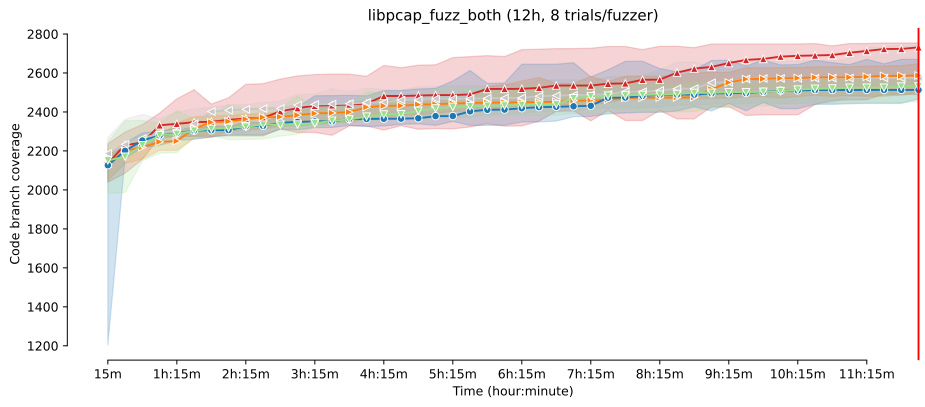}}
  \subfloat[Reached code coverage distribution
  on libpcap\_fuzz\_both]{\includegraphics[width = 0.3\textwidth]{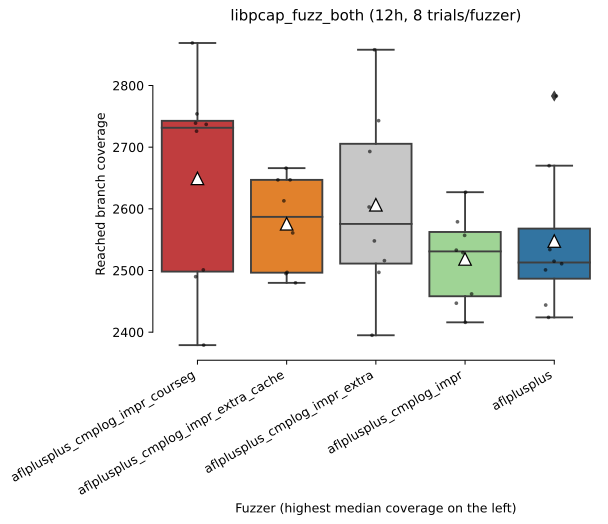}} \\

  \subfloat{\includegraphics[width = 0.9\textwidth]{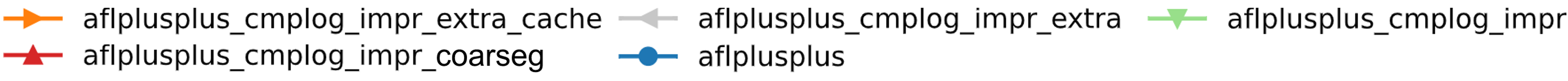}}
  \caption{Code coverage results of the improved variants compared to AFL++}
  \label{fig:variants}
\end{figure*}

\section{Evaluation}

In this section, the suggested improvements discussed in Section \ref{sec:improvements} are evaluated.

All benchmark trials are performed locally through the FuzzBench framework\cite{fuzzbench} on a private 64-bit KVM VM with 8 cores. Each trial is run for 12 hours on a single dedicated core\footnote{The complete fuzzbench report is available at \href{https://sanwieb.github.io/fuzzbench_reports/cmplog_improvement/}{https://sanwieb.github.io/fuzzbench\_reports/\\cmplog\_improvement/}}.

\subsection{Variants and targets}
For evaluation, the following variants were considered to demonstrate their relative improvements:

\begin{itemize}
  \item aflplusplus\_cmplog\_impr: AFL++ with the fine-grained improvement variant (Section \ref{sec:fine}).
  \item aflplusplus\_cmplog\_impr\_coarseg: AFL++ with the coarse-grained improvement variant (Section \ref{sec:coarse}).
  \item aflplusplus\_cmplog\_impr\_extra: AFL++ with the fine-grained improvement variant including the additional optimisations (Sections \ref{sec:fine}, \ref{sec:additional}).
  \item aflplusplus\_cmplog\_impr\_extra\_cache: AFL++ with the fine-grained improvement variant with the additional optimisations and the caching improvement (Sections \ref{sec:fine}, \ref{sec:additional}, \ref{sec:caching}).
\end{itemize}

Both the \textit{bloaty\_fuzz\_target} and the \textit{libpcap\_fuzz\_both} binaries were benchmarked, since these binaries showed a significant effect (negative and positive, respectively) on the performance with the original CmpLog feature enabled.

\subsubsection{Results}
\label{sec:varres}
\input{tables/stats_original.tex}

Figure \ref{fig:variants} shows the experimental code coverage results. A great improvement for \textit{bloaty\_fuzz\_target} (Figures \ref{fig:variants}(a), \ref{fig:variants}(b)) for all variants. Analysing Table \ref{tab:stats_original}, which shows the statistics of the trials, it must be noted that the coarse-grained variant has a larger maximum and average executions/time per seed. This follows our intuition since the fine-grained variant results in fewer magic byte candidates.

The largest boost is achieved when the additional improvements are enabled. The average number of executions per seed of `\textit{aflplusplus\_cmplog\_impr\_extra}' is more than halved (0.44) compared to the fine-grained improvement. Further improvement by enabling caching, reaches the highest code coverage and also has the steepest gain. Compared to the original AFL++, the cache variant reaches on average 30\% more code.

Another benchmark was performed on `\textit{libpcap\_fuzz\_both}' to ensure that the implementation did not negatively impact performance for which the original CmpLog already performed well. Even though the original CmpLog already performed well, all improvement variants reach on average more code coverage than the original implementation during the same period of benchmarking (Figures \ref{fig:variants}(c), \ref{fig:variants}(d)).

To conclude, the improved variants show a big potential for improvement of the CmpLog feature, especially for targets with similar roadblocks as \textit{`bloaty\_fuzz\_target'}.

\begin{figure*}
  \subfloat[Mean code coverage growth on bloaty\_fuzz\_target]{\includegraphics[width = 0.6\textwidth]{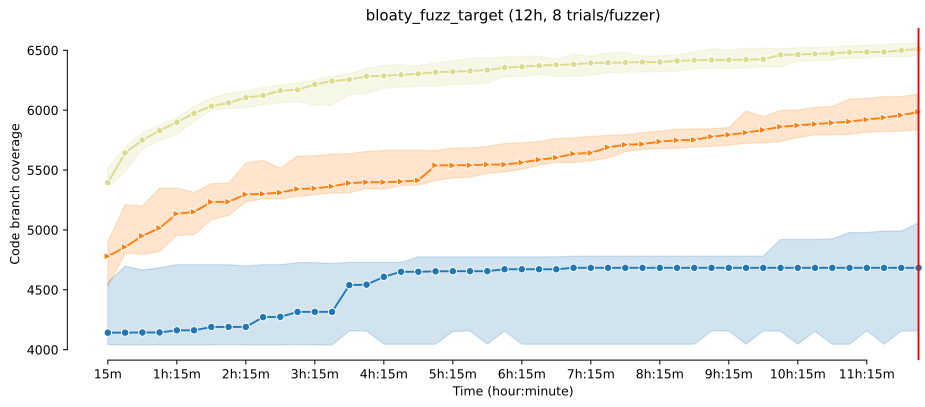}}
  \subfloat[Reached code coverage distribution on
  bloaty\_fuzz\_target]{\includegraphics[width = 0.3\textwidth]{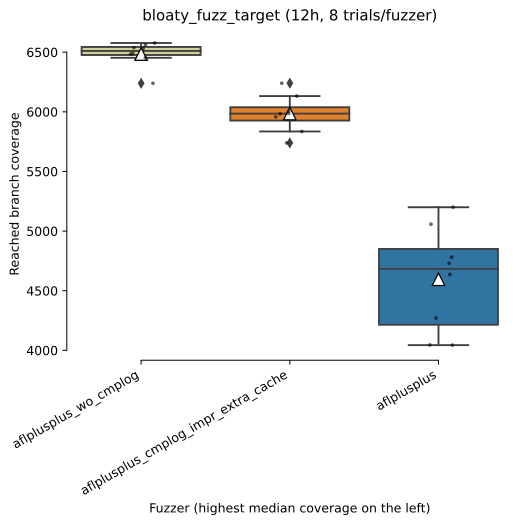}} \\

  \subfloat[Mean code coverage growth on libpcap\_fuzz\_both]{\includegraphics[width = 0.6\textwidth]{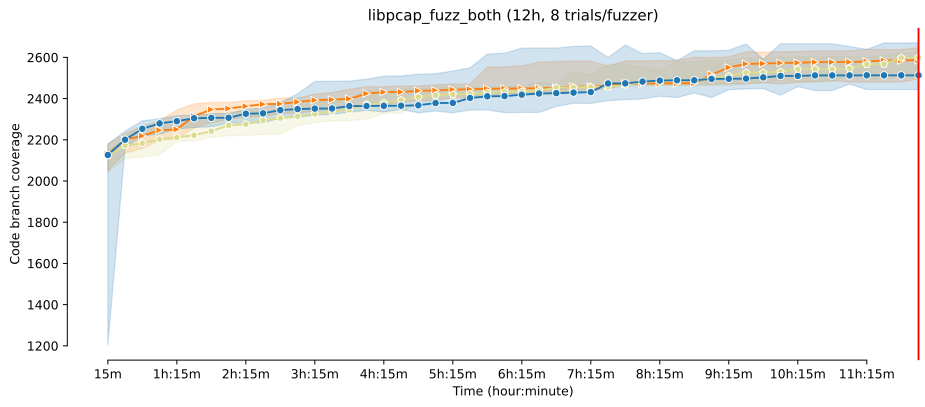}}
  \subfloat[Reached code coverage distribution on
  libpcap\_fuzz\_both]{\includegraphics[width = 0.3\textwidth]{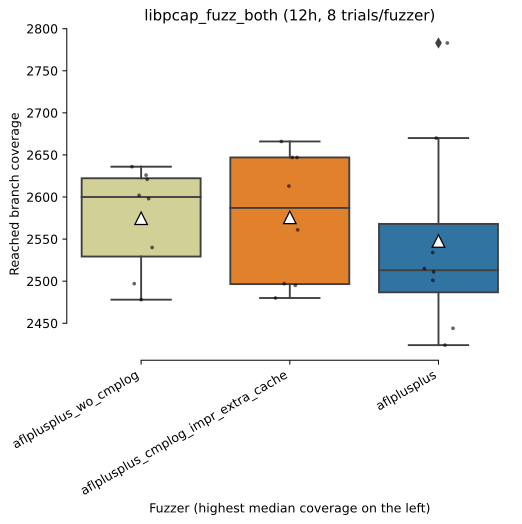}} \\

  \subfloat[Reached code coverage distribution on
  libpng-1.2.56]{\includegraphics[width = 0.6\textwidth]{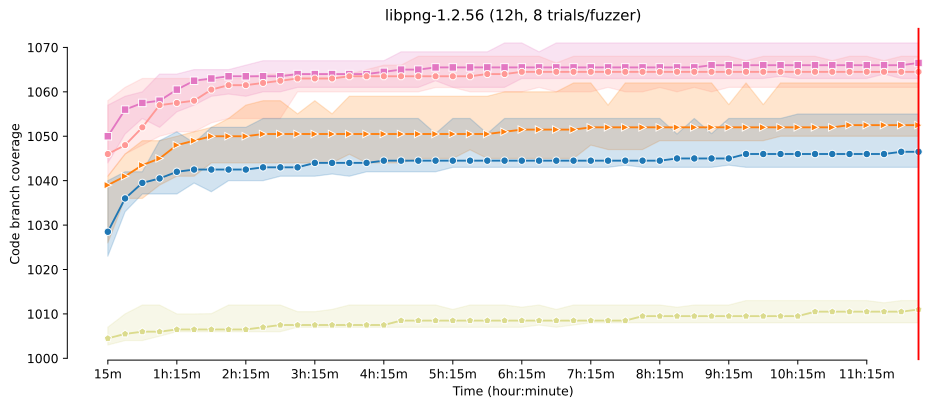}}
  \subfloat[Reached code coverage distribution on
  libpng-1.2.56]{\includegraphics[width = 0.3\textwidth]{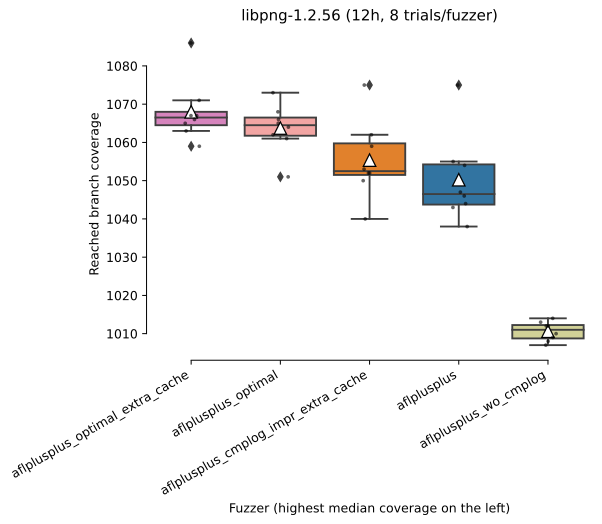}} \\
  
  \subfloat{\includegraphics[width = 0.9\textwidth]{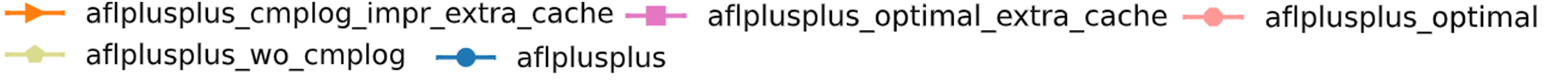}}
  \caption{Code coverage results of the extra cache variant compared to AFL++ default, AFL++ without CmpLog and AFL++ Optimal}
  \label{fig:compare}
\end{figure*}

\input{tables/stats_compare.tex}

\subsection{Comparison with the original AFL++}

Based on Section \ref{sec:varres}, `aflplusplus\_cmplog\_impr\_extra\_cache' was selected as the most promising set of improvements. In this section, this improvement is compared to the default AFL++ and AFL++ with CmpLog disabled. In addition to previous targets, the \textit{libpng-1.2.56} binary was benchmarked against the AFL++ optimal configuration for both the original CmpLog and the improved CmpLog implementation. The optimal configuration for \textit{libpng-1.2.56} is with CmpLog and the transformations and arithmetic operations enabled, as shown in Table \ref{tab:optimal}. Note that these two operations do not allow checking whether the pattern exists in the input leading to more executions which can impact the performance.

\subsubsection{Results}

Figure \ref{fig:compare} shows the code coverage results of the experiments. For the \textit{bloaty\_fuzz\_target}, both AFL++ with CmpLog disabled performs better, as well as the default AFLplusplus configuration. Analysing the statistics, Table \ref{tab:stats_cmp}, a significant difference in the execution speed can be observed; AFL++ with CmpLog disabled runs $2.6$ times more executions per second, partially explaining the performance difference. A further decrease of the introduced overhead by CmpLog might resolve this.

For \textit{`libpcap\_fuzz\_both'}, no significant performance difference were discerned between the improved version and the version without CmpLog.

For target \textit{libpng-1.2.56}, the non-optimal improved version reaches a higher code coverage than the default AFL++, and both optimal configurations outperform the default AFL++ and AFL++ without CmpLog. Comparing the two optimal configurations, the improved version reaches a slightly higher code coverage which is significant ($p < 0.5$). Moreover, looking at the statistics, a higher execution per second for the improved version is obtained. To conclude, the improved version performs well and even slightly outperforms the manually determined optimal variant.

\section{Future Work}

This research showed the potential of improving the original CmpLog feature implementation. Apart from the implemented improvements there is ample room for a deeper investigation.

\subsection{Generalize caching approach}
The improved cache variant showed the potential of using a caching mechanism. Making the implementation more generic by implemented a least-recently-used (LRU) cache can further boost performance. In this way, the most replaced bytes are cached instead of always the zeros and ones.

\subsection{Reconsider RedQueen implementation decisions}
The CmpLog differs from the original RedQueen in certain aspects. The biggest difference is the way of colourization. RedQueen tries to generate more entropy in the input bytes while the execution path remains the same \cite{aschermann2019redqueen}. CmpLog, however, replaces every byte with the same type. One of the consequences of this approach is discussed in Section \ref{sec:patterns} for the CMP instructions type; every 0 is replaced by a 1 which will result in a lot of possible candidates. The proposed improvement mostly tackles the resulting execution explosion problem, however, generating more entropy could further enhance it. Alternatively one could consider replacing multiple bytes in a row at once, thereby creating entropy with consecutive series.

Furthermore, RedQueen also includes a checksum identification mechanism which finds possible checksum checks and tries to pass them using temporary patching. CmpLog does not include any checksum identification. Considering the results of this research, this would be a valuable addition. The checksum identification can save many executions as an operand value controlled by a large input region is not I2S.

\subsection{Dynamically switching AFL++ features on or off}
This research showed that the cause of a feature not working for a certain binary can be very complex. For \textit{bloaty\_fuzz\_target} and CmpLog it was the combination of lots of comparisons, many taint bytes, and low entropy; factors which are not trivially detected through a pre-analysis script. However, while fuzzing, there are clear indications that CmpLog does not work for the target (execution explosion). One could consider developing a 'smart mode' in AFL++ to determines an optimal configuration for the PUT.

\section{Conclusion}

A deep dive into the CmpLog feature and its behaviour was performed, revealing some problems and design flaws hindering the overall performance. The proposed implementation improvement demonstrates the potential and promotes a redesign of the CmpLog implementation. Simultaneously, this research revealed the complexity of performance problems for a target-fuzzer combination. The improvement CmpLog version is made available on \href{https://github.com/SanWieb/AFLplusplus/tree/cmplog_improvement_stable}{https://github.com/SanWieb/AFLplusplus/tree/cmplog\_improvement\_stable}

\printbibliography

\end{document}

%% file: tables/fuzzbench_all.tex
\begin{table}[t]
    \centering
    \small
    \begin{tabular}{lllll}
    \hline
                          & AFL\cite{zalewski2017american} & AFLFast\cite{bohme2016coverage}        & libfuzzer\cite{serebryany2015libfuzzer} & AFL++\cite{fioraldi2020afl++}                       \\ \hline
    openthread-2018-02-27 & 6208                        & {\color{red} 5549}          & 5957      & \textbf{7145}               \\
    libpng-1-2.56         & {\color{red} 1944}          & {\color{red} 1942}          & 1995      & \textbf{2089}               \\
    vorbis 2017-12-11     & 2167                        & 2161                        & 1928      & \textbf{2172}               \\
    bloaty\_fuzz\_target           & \textbf{8278}               & 8008                        & 7051      & {\color{red} 6466} \\
    harfbuzz              & 8407                        & 8361                        & 8334      & \textbf{8715}               \\
    libpcap\_fuzz\_both               & {\color{red} 98}            & {\color{red} 95}            & 3461      & \textbf{4331}               \\
    zlib\_zlib            & 961                         & 944                         & 975       & 963                         \\ \hline
    \end{tabular}
    \caption{Fuzzbench 2022-04-19 code-coverage results. (bold text = relative high code coverage or fastest gain, red = relative low code coverage)}
    \label{tab:fuzzbench_all}
\end{table}

%% file: tables/optimal.tex
\begin{table*}[t]
    \small
    \centering
    \resizebox{\textwidth}{!}{\begin{tabular}{lllll}
    \hline
                          & AFL++ & AFL++ Opt. & Build config Opt.          & Run config Opt.                            \\ \hline
    openthread-2018-02-27 & 6849  & 6736       & Default (LTO, Cmplog)      & Cmplog arthemetic; Keep Timeouts           \\
    libpng-1-2.56         & 2091  & 2122       & Default (LTO, Cmplog)      & Cmplog transformations \& arthemetic; Keep Timeouts           \\
    vorbis 2017-12-11     & 2173  & 2180       & LTO, LAF                   & Test cache size 50                         \\
    bloaty fuzz           & 6553  & 8767       & LTO                        & MOpt immediatly                            \\
    harfbuzz              & 8721  & 8654       & Cmplog, Dict2File, Tracepc & Keep timeouts                              \\
    libpcap               & 4444  & 4559       & LTO, Cmplog                & Cmplog transformations, Test cache size 50 \\
    zlib\_zlib            & 967   & 967        & Cmplog, Dict2File, Tractpc &                                            \\ \hline
    \end{tabular}}
    \caption[Caption for LOF]{AFL++ optimal configuration compared to the default configuration\textsuperscript{\ref{benchmarkgoogle}.}}
    \label{tab:optimal}
\end{table*}

%% file: tables/coverage.tex
\begin{table}[t]
    \centering
    \small
    \begin{tabular}{llll}
    \hline
                                        & AFL++                           & AFL++ Optimal                   & Initial Seeds                   \\ \hline
    \multicolumn{1}{l|}{bloaty.cc}      & \cellcolor[HTML]{FFD0D0}33.23\% & \cellcolor[HTML]{FFD0D0}33.23\% & \cellcolor[HTML]{FFD0D0}31.17\% \\
    \multicolumn{1}{l|}{bloaty.h}       & \cellcolor[HTML]{FFFFD0}87.50\% & \cellcolor[HTML]{FFFFD0}87.50\% & \cellcolor[HTML]{FFD0D0}62.50\% \\
    \multicolumn{1}{l|}{demangle.cc}    & \cellcolor[HTML]{FFD0D0}47.01\% & \cellcolor[HTML]{FFFFD0}90.92\% & \cellcolor[HTML]{FFD0D0}1.37\%  \\
    \multicolumn{1}{l|}{disassemble.cc} & \cellcolor[HTML]{FFD0D0}17.54\% & \cellcolor[HTML]{FFD0D0}17.54\% & \cellcolor[HTML]{FFD0D0}16.67\% \\
    \multicolumn{1}{l|}{dwarf.cc}       & \cellcolor[HTML]{FFFFD0}92.54\% & \cellcolor[HTML]{FFFFD0}94.96\% & \cellcolor[HTML]{FFD0D0}71.02\% \\
    \multicolumn{1}{l|}{elf.cc}         & \cellcolor[HTML]{FFFFD0}91.09\% & \cellcolor[HTML]{FFFFD0}91.09\% & \cellcolor[HTML]{FFFFD0}87.74\% \\
    \multicolumn{1}{l|}{macho.cc}       & \cellcolor[HTML]{FFD0D0}62.15\% & \cellcolor[HTML]{FFFFD0}88.55\% & \cellcolor[HTML]{FFD0D0}29.38\% \\
    \multicolumn{1}{l|}{range\_map.cc}  & \cellcolor[HTML]{FFFFD0}86.89\% & \cellcolor[HTML]{FFFFD0}86.89\% & \cellcolor[HTML]{FFFFD0}80.33\% \\
    \multicolumn{1}{l|}{range\_map.h}   & \cellcolor[HTML]{FFD0D0}54.55\% & \cellcolor[HTML]{FFD0D0}54.55\% & \cellcolor[HTML]{FFD0D0}77.42\% \\
    \multicolumn{1}{l|}{webassembly.cc} & \cellcolor[HTML]{FFD0D0}77.91\% & \cellcolor[HTML]{FFFFD0}82.56\% & \cellcolor[HTML]{FFD0D0}34.62\% \\ \hline
    Total:                              & 7.22\%                          & 8.76\%                          & 4.73\%                          \\ \hline
    \end{tabular}
    \caption{Branch coverage reached by AFL++, AFL++ Optimal and-
    the coverage with the initial seeds (red / yellow colored: below / above 80\%). Shared libraries are excluded from the table, but included in the total calculation.}
    \label{tab:coverage}
\end{table}

%% file: tables/stats_default_optimal.tex
\begin{table}[t]
    \small
    \centering
    \begin{tabular}{llll}
    \hline
                             & AFL++       & AFL++ Optimal                    & Ratio \\ \hline
    \textbf{execs\_done}     & 198376360.6 & \multicolumn{1}{l|}{388016450.6} & 1.96     \\
    \textbf{execs\_per\_sec} & 2396.4      & \multicolumn{1}{l|}{4687.55}     & 1.96     \\
    \textbf{cycles\_done}    & 0           & \multicolumn{1}{l|}{31.45}       & -        \\
    \textbf{corpus\_count}   & 1581.85     & \multicolumn{1}{l|}{6708}        & 4.24     \\
    \textbf{corpus\_favored} & 379.7       & \multicolumn{1}{l|}{1033.6}      & 2.72     \\ \hline
    \end{tabular}
    \caption{Average statistics for the trials of AFL++ (trials=18) and AFL++ Optimal (trials=20).}
    \label{tab:stats_default_optimal}
\end{table}

%% file: tables/seed_orig.tex
\begin{table}[t]
    \small
    \centering
    \begin{tabular}{lll}
    \hline
                            & AFL++   & AFL++ Optimal \\ \hline
    \textbf{Avg time}       & 53 min  & 2 min        \\
    \textbf{Max time}       & 199 min & 13 min        \\
    \textbf{Avg Executions} & 7633 k  & 425 k         \\
    \textbf{Max Executions} & 22742 k & 3395 k       \\ \hline
    \end{tabular}
    \caption{Average time and number of executions per seed for the top 15 seeds, information retrieved from the plot\_data files of AFL++ (trials=18) and AFL++ Optimal (trials=20).}
    \label{tab:seed_orig}
\end{table}


%% file: tables/rtn_operand.tex
\begin{table*}[t]
    \small
    \centering
    \resizebox{\textwidth}{!}{\begin{tabular}{llllll}
    \hline
    \#         & Type      & Length    & Executions &               &                                              \\ \hline
    \textbf{1} & RTN       & 31        & 28         & Colorized o0: & \texttt{388400c4127f0000388400c4127f0000308400c4127f0000202d00c4127f00} \\
               & \textbf{} & \textbf{} &            & Original o0:  & \texttt{786800c4127f0000786800c4127f0000706800c4127f0000803500c4127f00} \\
    \textbf{2} & RTN       & 31        & 42421      & Colorized o1: & \texttt{000000000000000018760048ec7f0000900d0048ec7f0000f84e4988655500} \\
               & \textbf{} & \textbf{} &            & Original o1:  & \texttt{0000000000000000687a0048ec7f0000c0590048ec7f000038dc4988655500} \\
    \textbf{3} & RTN       & 31        & 84487      & Colorized o0: & \texttt{000000000000000000000000000000000000000000000000800f0048ec7f00} \\
               &           &           &            & Original o0:  & \texttt{000000000000000000000000000000000000000000000000900d0048ec7f00} \\ \hline
    \end{tabular}}
    \caption{Examples of operand values of RTN comparisons with the number of produced executions, showed is either the left or right operand.}
    \label{tab:rtn-operand}
\end{table*}


%% file: tables/ins_operand.tex
\begin{table*}[t]
    \small
    \centering
    \begin{tabular}{llllllll}
    \hline
    \#         & Type & Length & Colorized o0           & Colorized o1 & Original o0 & Original o1 & Executions \\ \hline
    \textbf{1} & INS  & 8      & 120 (2001)             & e8 (e800)    & 640 (4006)  & 614 (1406)  & 0          \\
    \textbf{2} & INS  & 8      & 1c8 (c801)             & 190 (9001)   & 40 (4000)   & 00 (00)     & 7087       \\
    \textbf{3} & INS  & 4      & 1010111 (1101)         & 08 (0800)    & 0e (0e00)   & 08(0800)    & 20400      \\
    \textbf{4} & INS  & 8      & 101010101010101 (0101) & 00 (00)      & 01 (100)    & 00 (00)     & 38668      \\ \hline
    \end{tabular}
    \caption{Examples of operand values of INS comparisons with the number of produced executions, showed is both the left and right operand.}
    \label{tab:ins-operand}
\end{table*}

%% file: tables/stats_original.tex
\begin{table*}[t]
    \small
    \centering
    \resizebox{\textwidth}{!}{
    \begin{tabular}{l|llllllll}
    \hline
    Fuzzer                                  & Binary               & \begin{tabular}[c]{@{}l@{}}cycles\\ done\end{tabular} & \begin{tabular}[c]{@{}l@{}}execs /\\ sec\end{tabular} & \begin{tabular}[c]{@{}l@{}}corpus\\ count\end{tabular} & \begin{tabular}[c]{@{}l@{}}MaxTime\\ / seed\end{tabular} & \begin{tabular}[c]{@{}l@{}}Avg Time\\ / seed\end{tabular} & \begin{tabular}[c]{@{}l@{}}Max \\ Executions\\ / seed\end{tabular} & \begin{tabular}[c]{@{}l@{}}Avg \\ Executions \\ / seed\end{tabular} \\ \hline
    aflplusplus                             & bloaty\_fuzz\_target & 0.00                                                  & 827.75                                                & 959.00                                                 & 288.58                                                   & 65.03                                                     & 13660                                                              & 3224                                                                \\
    aflplusplus\_cmplog\_impr               & bloaty\_fuzz\_target & 0.00                                                  & 869.67                                                & 1684.92                                                & 119.30                                                   & 44.91                                                     & 5835                                                               & 2250                                                                \\
    aflplusplus\_cmplog\_impr\_coarseg      & bloaty\_fuzz\_target & 0.00                                                  & 856.62                                                & 1541.62                                                & 197.03                                                   & 53.18                                                     & 9052                                                               & 2608                                                                \\
    aflplusplus\_cmplog\_impr\_extra        & bloaty\_fuzz\_target & 0.00                                                  & 836.36                                                & 2509.21                                                & 32.12                                                    & 20.29                                                     & 1596                                                               & 1001                                                                \\
    aflplusplus\_cmplog\_impr\_extra\_cache & bloaty\_fuzz\_target & 0.50                                                  & 803.33                                                & 3255.83                                                & 22.76                                                    & 13.40                                                     & 1169                                                               & 649                                                                 \\ \hline
    aflplusplus                             & libpcap\_fuzz\_both  & 40.75                                                 & 5330.25                                               & 1758.12                                                & 54.58                                                    & 7.83                                                      & 17712                                                              & 2529                                                                \\
    aflplusplus\_cmplog\_impr               & libpcap\_fuzz\_both  & 43.12                                                 & 5527.75                                               & 1750.25                                                & 72.21                                                    & 9.09                                                      & 24085                                                              & 3031                                                                \\
    aflplusplus\_cmplog\_impr\_coarseg      & libpcap\_fuzz\_both  & 37.75                                                 & 5322.62                                               & 1925.00                                                & 49.85                                                    & 7.18                                                      & 15996                                                              & 2299                                                                \\
    aflplusplus\_cmplog\_impr\_extra        & libpcap\_fuzz\_both  & 35.50                                                 & 5354.62                                               & 1852.38                                                & 79.03                                                    & 10.38                                                     & 25681                                                              & 3364                                                                \\
    aflplusplus\_cmplog\_impr\_extra\_cache & libpcap\_fuzz\_both  & 41.25                                                 & 5302.62                                               & 1661.25                                                & 56.70                                                    & 8.75                                                      & 18096                                                              & 2811                                                                \\ \hline
    \end{tabular}}
    \caption{Average trial statistics of the improved variants compared to AFL++ (8 trials/fuzzer).}
    \label{tab:stats_original}
\end{table*}

%% file: tables/stats_compare.tex
\begin{table*}[t]
    \small
    \centering
    \resizebox{\textwidth}{!}{\begin{tabular}{l|llllllll}
    \hline
    Fuzzer                                  & Binary               & \begin{tabular}[c]{@{}l@{}}cycles\\ done\end{tabular} & \begin{tabular}[c]{@{}l@{}}execs /\\ sec\end{tabular} & \begin{tabular}[c]{@{}l@{}}corpus\\ count\end{tabular} & \begin{tabular}[c]{@{}l@{}}MaxTime\\ / seed\end{tabular} & \begin{tabular}[c]{@{}l@{}}Avg Time\\ / seed\end{tabular} & \begin{tabular}[c]{@{}l@{}}Max \\ Executions\\ / seed\end{tabular} & \begin{tabular}[c]{@{}l@{}}Avg \\ Executions \\ / seed\end{tabular} \\ \hline
    aflplusplus                             & bloaty\_fuzz\_target & 0.00                                                  & 827.75                                                & 959.00                                                 & 288.58                                                   & 65.03                                                     & 13660                                                              & 3224                                                                \\
    aflplusplus\_cmplog\_impr\_extra\_cache & bloaty\_fuzz\_target & 0.50                                                  & 803.33                                                & 3255.83                                                & 22.76                                                    & 13.40                                                     & 1169                                                               & 649                                                                 \\
    aflplusplus\_wo\_cmplog                 & bloaty\_fuzz\_target & 9.12                                                  & 2094.50                                               & 5726.50                                                & 14.64                                                    & 2.62                                                      & 1899                                                               & 312                                                                 \\ \hline
    aflplusplus                             & libpcap\_fuzz\_both  & 40.75                                                 & 5330.25                                               & 1758.12                                                & 54.58                                                    & 7.83                                                      & 17712                                                              & 2529                                                                \\
    aflplusplus\_cmplog\_impr\_extra\_cache & libpcap\_fuzz\_both  & 41.25                                                 & 5302.62                                               & 1661.25                                                & 56.70                                                    & 8.75                                                      & 18096                                                              & 2811                                                                \\
    aflplusplus\_wo\_cmplog                 & libpcap\_fuzz\_both  & 42.00                                                 & 5572.75                                               & 1766.50                                                & 57.61                                                    & 7.73                                                      & 19521                                                              & 2614                                                                \\ \hline
    aflplusplus                             & libpng-1.2.56        & 167.88                                                & 20790.12                                              & 759.25                                                 & 66.75                                                    & 11.25                                                     & 84935                                                              & 14050                                                               \\
    aflplusplus\_cmplog\_impr\_extra\_cache & libpng-1.2.56        & 164.12                                                & 20822.75                                              & 782.38                                                 & 68.82                                                    & 11.80                                                     & 86332                                                              & 14854                                                               \\
    aflplusplus\_wo\_cmplog                 & libpng-1.2.56        & 155.75                                                & 8871.50                                               & 693.50                                                 & 72.69                                                    & 11.19                                                     & 61946                                                              & 9787                                                                \\
    aflplusplus\_optimal                    & libpng-1.2.56        & 175.25                                                & 17910.38                                              & 783.50                                                 & 53.20                                                    & 9.03                                                      & 57544                                                              & 9670                                                                \\
    aflplusplus\_optimal\_extra\_cache      & libpng-1.2.56        & 169.50                                                & 21900.12                                              & 809.12                                                 & 46.84                                                    & 8.79                                                      & 61681                                                              & 11656                                                               \\ \hline
    \end{tabular}}
    \caption{Average trial statistics of the extra cache variant with different AFL++ configurations (8 trials/fuzzer).}
    \label{tab:stats_cmp}
\end{table*}